\documentstyle[aps,prl,psfig,floats]{revtex}
\begin{document}
%\draft

\twocolumn[\hsize\textwidth\columnwidth\hsize\csname @twocolumnfalse\endcsname

\title{Stochastic Ballistic Annihilation and Coalescence\\}

\author{R. A. Blythe$^{(1)}$, M. R. Evans$^{(1)}$, Y. Kafri$^{(2)}$}

\address{$^{1}$ Department of Physics and Astronomy, University of
        Edinburgh, Mayfield Road, Edinburgh EH9 3JZ, U.K.\\ $^{2}$
        Department of Physics of Complex Systems, The Weizmann Institute 
        of Science, Rehovot 76100, Israel\\[-2mm] $ $}
        
\date{May 3rd 2000; revised Aug 4th 2000}

\maketitle
\begin{abstract}
  We study a class of stochastic ballistic annihilation and
  coalescence models with a binary velocity distribution in one
  dimension.  We obtain an exact solution for the density which
  reveals a universal phase diagram for the asymptotic density decay.
  By universal we mean that all models in the class are described by a
  single phase diagram spanned by two reduced parameters.  The phase
  diagram reveals four regimes, two of which contain the previously
  studied cases of ballistic annihilation. The two new phases are a
  direct consequence of the stochasticity.  The solution is obtained
  through a matrix product approach and builds on properties of a
  $q$-deformed harmonic oscillator algebra.
\end{abstract}

\vspace{2mm}
]
%%%%%%%%%%%%%%%%%%%%%%%%%%%%%%%%%%%%%%%%%%%%%%%%%%%%%%%%%%%%%%%%%%%%%%%%%%%%%
% explain PACS:                                                             %
%-------------                                                              %
%PACs   05.40.-a Fluctuation phenomena                                      %
%       02.50.Ey Stochastic processes                                       %
%       82.20.Mj Nonequilibrium kinetics                                    %
%                                                                           %
%%%%%%%%%%%%%%%%%%%%%%%%%%%%%%%%%%%%%%%%%%%%%%%%%%%%%%%%%%%%%%%%%%%%%%%%%%%%%
%                                 MAIN TEXT                                 %
%%%%%%%%%%%%%%%%%%%%%%%%%%%%%%%%%%%%%%%%%%%%%%%%%%%%%%%%%%%%%%%%%%%%%%%%%%%%%

Systems of reacting particles are used to model a whole gamut of
phenomena relevant to fields ranging from chemical physics through
statistical physics to mathematical biology.  In some applications the
particles represent chemical or biological species \cite{Kopelman,HZ};
in other cases they are to be interpreted as composite objects such as
aggregating traffic jams \cite{BKS}.  Excitations can also be treated
as interacting particles, one example being laser-induced excitons in
certain crystals \cite{exp}.  Furthermore, domain walls occurring in a
number of different contexts such as growth and coarsening processes
\cite{KS,GK} have dynamics with a natural particle interpretation.

Generally these systems are defined through nonequilibrium dynamics.
Given such a wide variety of nonequilibrium reaction systems, it is
natural to ask if they can be divided into distinct groups akin to
the universality classes known for equilibrium systems.

Two reactions that have been extensively studied are single species
annihilation ($A{+}A{\to}\emptyset$) and coalescence ($A{+}A{\to}A$).
A particularly striking result is that if the reactant motion is
diffusive, the two processes belong to the same universality class
\cite{diffuse} and the density decay is independent of the reaction
rate in two dimensions and below.  Moreover, these diffusive systems
have also served as prototypes for the development of a variety of
theoretical tools ranging from field theoretic renormalization group
(RG) \cite{Cardy}, to exact methods in low dimensions \cite{Privman}.

On the other hand, much less is known about the same reactions when
the motion is ballistic (deterministic) despite the relevance of such
motion to the modeling of growth and coarsening processes
\cite{KS,GK}.  A seminal model was introduced and solved by Elskens
and Frisch \cite{EF} and describes pairwise annihilation of oppositely
moving particles in one dimension (1D).  That study was restricted to
particles that react upon contact with probability one.  More recently
some results have been obtained for systems in which the reaction
probability is less than one, thus introducing stochasticity into the
evolution \cite{Richardson,Kafri}.

In this work we introduce a class of 1D stochastic
ballistic reaction systems.  The class includes both ballistic
annihilation and coalescence and incorporates as special limits the
models of \cite{EF} and \cite{Kafri}.  We obtain an exact solution for
the density decay which reveals a single phase diagram common to all
combinations of ballistic annihilation and coalescence. This
demonstrates a universality of the two processes.  The universality is
stronger than that usually discussed in an RG context and can be
likened to a law of corresponding states.  The phase diagram generically
comprises four decay regimes in contrast to the two previously known
\cite{EF}.  The new phases are a result of the stochasticity
of the reactions. 

Our exact solution is based on the invariance of certain properties of
our class of models under change of the initial spacing of the
particles. As a consequence the long time density may be determined
exactly through a matrix product approach of the type introduced in
\cite{DEHP}.  We use this property, and employ recent results on
$q$-deformed algebras \cite{BECE}, to analyse the asymptotic density
decay.

\begin{figure}
\center{\psfig{figure=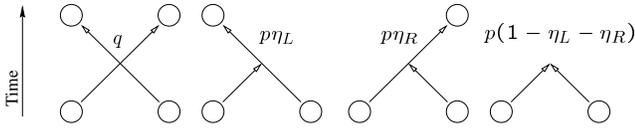,width=8.4cm}}
\caption{\label{fig:outcomes}The possible reactions and their probabilities.}
\end{figure}

We now define the class of models to be considered.  At time $t=0$
reactants are placed on a line with nearest-neighbor distances chosen
independently from a continuous exponential distribution.  The unit of
length is chosen so that the initial density is $\varrho=1$.  Although
we consider here a Poisson initial condition, our methods are
extendible to more general initial distributions (see below).  Each
particle is assigned a velocity $+c$ (right-moving) or $-c$
(left-moving) with probability $f_R$ and $f_L = 1 - f_R$ respectively.
Particles move ballistically until two collide, at which point one of
four outcomes follows, see Fig.~\ref{fig:outcomes}: the particles pass
through each other with probability $q$; the particles coalesce into a
left (right) moving particle with probability $p \eta_L$ ($p \eta_R$);
the particles annihilate with probability $p(1-\eta_L-\eta_R)$.  Here
$p=1-q$ is the probability that some reaction occurs.

\begin{figure}
\center{\psfig{figure=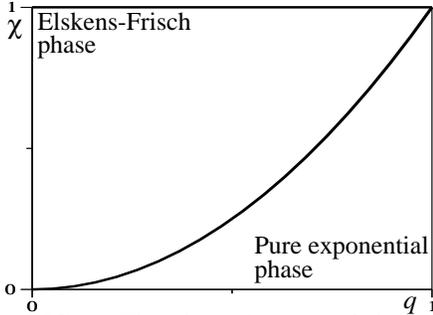,width=5.8cm}}
\caption{\label{fig:phasediag}The phase diagram of the model for $\chi \le 1$.
The behavior for
  $\chi > 1$ can be deduced from that with $\chi < 1$ via a left-right
  symmetry (see text).}
\end{figure}

Before describing our method of solution, we present our main results
summarized in the phase diagram Fig.~\ref{fig:phasediag}.  Two
important quantities that emerge are the reduced densities $f_L^\ast =
f_L (1-\eta_R)$ and $f_R^\ast = f_R (1-\eta_L)$.  The phase diagram is
spanned by $q$, the probability of not reacting, and $\chi= f_R^\ast/
f_L^\ast $, the ratio of reduced densities.  For simplicity we
consider $\chi \leq 1$ which results in eventual extinction of
right-moving particles; $\chi=1$ is a special case where both species
die out.  The case $\chi >1$ can be treated using the symmetry of the
model under left-right interchange $f_R \leftrightarrow f_L$
and $\eta_R \leftrightarrow \eta_L$ i.e.\ $\chi \to 1/\chi$.

The different phases---of which two are regions and two are lines in
the phase diagram---correspond to four qualitative long-time density
decays.  When $\chi < q^2$ the decay is purely exponential of the form
$\varrho \simeq \varrho_\infty + b_1 \exp(- \alpha_1 t)$; when $\chi =
q^2$ the decay is exponential, multiplied by a $t^{-1/2}$ power law
$\varrho \simeq \varrho_\infty + b_2 \exp(- \alpha_2 t)/t^{1/2}$; when
$q^2{ < }\chi{ < }1$ the decay exponential, multiplied by a $t^{-3/2}$
power law $\varrho \simeq \varrho_\infty + b_3 \exp(- \alpha_3
t)/t^{3/2}$; finally when $\chi=1$ the decay is pure power law
$\varrho \simeq b_2/t^{1/2}$.  The exact expressions for the
coefficients $\alpha,b$ and final density $\varrho_\infty$ are as
given in Table~\ref{table:phases}.  Some intriguing features that
emerge from the phase diagram Figure~\ref{fig:phasediag} are:

({\it i}) There is universality of ballistic annihilation and
coalescence. This is manifested by the fact that all the information
concerning the reactions of a particular model, along with the initial
densities, are encoded into a single parameter $\chi$.  For a
generic choice of $\eta_R$, $\eta_L$ defining a particular
annihilation-coalescence model, the same four decay regimes are found
by varying the initial densities or stochasticity parameter $q$.  In
this way the universality can be considered as a law of corresponding
states. 

({\it ii}) Two new density decays appear which were not anticipated in
previous works.  The first is the line $\chi=q^2$ and the second the
region $\chi < q^2$.  Thus for a generic value of the reaction
probability $1-q$, varying the initial densities gives rise to four
types of asymptotic decay.

({\it iii}) The deterministic case $q=0$ is non-generic since along
this line only two of the possible phases are traversed.  For the pure
annihilation model ($\eta_R=\eta_L=0$) these phases were found in
\cite{EF}. Thus we refer to the entire region $q^2{}<{}\chi{}<1$ as
the Elskens-Frisch phase.

({\it iv}) The line $\chi =1$ (equal reduced densities) is non-generic
since a single, power law, decay regime is found. The decay does not
depend on the stochasticity $q$. For $\eta_R=\eta_L=0$ this special
line corresponds to equal initial densities \cite{Kafri}. Our results
show that such a special line exists for all combinations of
annihilation and coalescence.  This phase can be understood through
the picture of \cite{EF}.  Density fluctuations in the initial
conditions lead to trains of left- and right-moving particles: in a
length $\sim t$ the excess particle number is $\sim t^{1/2}$ which
yields the $t^{-1/2}$ density decay.  At long times, the train size is
large and so a particle in one train encounters many particles in the
other and will eventually react making the parameter $q$ irrelevant.

({\it v}) In the two new phases ($\chi \le q^2$) the two species decay
at {\it unequal} rates leaving a non-zero population of left-moving
particles.  This is to be contrasted to the Elskens-Frisch phase
($\chi > q^2$) where the final density of one species is non-zero but
both species decay at the same rate.  A simple example of non-equal
decays is the case $\eta_R=0, \eta_L=1$ ($\chi =0$).  Then
left-moving particles do not decay but simply absorb the right-moving
particles with probability $1-q$ giving $\varrho = f_R {\rm e}^{-2
  (1-q) f_L c t}$.  Our results show that, in general, increasing $q$
leads to a non-trivial transition at $\chi=q^2$ to a regime where the
two species have different decay forms.
\begin{table}[htb]
\begin{center}
\begin{tabular}{c|c}
 & $\varrho(t)-\varrho_\infty$\\[0.5ex]
\hline &\\
$ \chi < q^2 $ &
$\displaystyle f_R \left( 1 - \frac{f_R^\ast}{q^2 f_L^\ast} \right) 
{\rm e}^{
 - 2\, \left(1-q\right) \left( f_L^\ast - f_R^\ast/q \right) ct }$
\\[2.9ex]
$ \chi = q^2 $ &
$\displaystyle \frac{f_R}{\sqrt{2\pi}} \left( \frac{1}{f_L^\ast f_R^\ast}
  \right)^{\!\frac{1}{4}} \frac{{\rm e}^{-2(\sqrt{f_L^\ast} - \sqrt{f_R^\ast})^2 c t}}
  {(ct)^{1/2}} $ 
\\[2.9ex]
$ q^2 {<} \chi {<} 1 $ & 
$\displaystyle \frac{K}{4\sqrt{2\pi}}
 \left( \frac{1}{f_L^\ast f_R^\ast}
 \right)^{\!\frac{3}{4}}
 \frac{f_L f_R^\ast + f_R f_L^\ast}{( \sqrt{f_L^\ast} -
   \sqrt{f_R^\ast})^2}
 \frac{{\rm e}^{-2 (\sqrt{f_L^\ast} - \sqrt{f_R^\ast})^2 ct}}{(ct)^{3/2}}
$
\\[2.9ex]
$ \chi = 1 $ &
$\displaystyle \frac{1}{\sqrt{2\pi}} \left( \frac{1}{f_L^\ast f_R^\ast}
  \right)^{\!\frac{1}{4}} \left(\frac{1}{ct} \right)^{\!\frac{1}{2}}$ 
\\[2.9ex]
\end{tabular}
\end{center}
\caption{\label{table:phases} Long time density decays
to the asymptotic value $\varrho_\infty=
f_L(1-\chi)$. Results for $\chi>1$ can be obtained via
the right-left symmetry (see text).
$K$ is given by (\ref{eqn:Kdef}).
}
\end{table}

We now turn to the method of derivation of the phase diagram.  The
density is given by
\begin{equation}
\label{eqn:rhodef}
  \varrho(t) = f_R  P_S^{(R)}(t) + f_L P_S^{(L)}(t) 
\end{equation}
where $P_S^{(L)}(t)$ and $P_S^{(R)}(t)$ are the probabilities that a
left- and right-moving particle survive up to a time $t$ respectively.
These two probabilities are related by the left-right symmetry noted
previously; therefore, if we calculate $P_S^{(R)}(t)$ for all $\chi$
we can infer $P_S^{(L)}(t)$.
\begin{figure}
\center{\psfig{figure=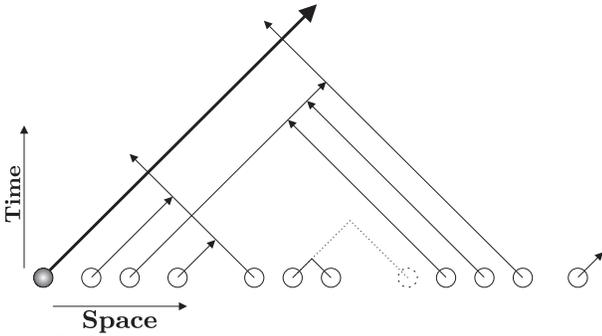,width=8cm}}
\caption{\label{fig:traj} 
  A configuration and set of trajectories and reactions for a
  test particle (shaded and bold line) encountering a string of $N=10$
  particles.  Note how changing the spacing between, for example, the
  fifth and sixth particles (indicated by dotted lines) alters the
  time sequence of the reactions but not the final survival
  probability.}
\end{figure}
To calculate $P_S^{(R)}(t)$ consider the leftmost, right-moving
particle in Fig.~\ref{fig:traj} which we refer to as a test particle.
From the figure one can see that the initial spacing of the particles
on the line does not affect the sequence of possible reactions for any
given particle, in particular for the test particle (we return to this
point later).  Also note that after a given time $t$, the test
particle may only have interacted with the $N$ particles initially
placed within a distance $X=2ct$ (and to the right) of the chosen
particle.  These two facts imply that the survival probability can be
expressed in terms of two independent functions. The first is $F(N)$,
the probability that the test particle survives reactions with the $N$
particles initially to its right, and depends only on the sequence of
the $N$ particles. The second is $G(N;X)$, the probability that
initially there were exactly $N$ particles in a region of size
$X=2ct$. Explicitly,
\begin{equation}
\label{eqn:Psum}
  P_S^{(R)}(X) = \sum_{N=0}^{\infty} F(N) G(N;X)\;.
\end{equation}
Thus the problem is reduced to two separate combinatorial problems
of calculating $G(N;X)$ and $F(N)$.  For the Poisson initial
conditions $G(N;X) = X^N \mbox{e}^{-X}/N!$ \cite{Feller}.  In the
following we show how the second problem may be solved by employing a
matrix product approach \cite{DEHP}.

As an example consider a test particle encountering the string of
reactants depicted in Fig.~\ref{fig:traj}. We claim that the
probability of the test particle surviving through this string may be
written as
\begin{equation}
\langle W | RRRLRLLLLR | V \rangle
\label{eqn:Fex}
\end{equation}
where $R,L$ are matrices (or operators) and $\langle W |$, $| V
\rangle$ vectors with scalar product $\langle W | V \rangle = 1$.
Thus we write, in order, a matrix $R$ ($L$) for each right (left)
moving particle in the initial string.

We now show that the conditions for an expression such as
(\ref{eqn:Fex}) to hold for an arbitrary string are
\begin{eqnarray}
\label{eqn:RL}
RL  &=& qLR + p(\eta_L L + \eta_R R + [1-\eta_L -\eta_R]) \\\label{eqn:VWdef}
  \langle W| L &=&  \langle W| (q+p\eta_R) \qquad
  R |V\rangle = |V\rangle\;.
\end{eqnarray}
To understand condition (\ref{eqn:RL}), recall that after an
interaction between a right-moving and left-moving particle there are
four possible outcomes (see Fig.~1) corresponding to the four terms on
the right hand side of Eq.~\ref{eqn:RL} with probabilities given by
the respective coefficients. Using (\ref{eqn:RL}), any initial matrix
product such as (\ref{eqn:Fex}) can be reduced to a sum of terms of
form $\langle W | L^sR^t | V \rangle$ corresponding to all possible
final states ensuing from the initial string and with coefficients
equal to the probabilities of each final state. The test particle will
survive such a final state and pass through the $s$ left-moving
particles with probability $(q+p\eta_R)^s$. The conditions
(\ref{eqn:VWdef}) ensure that this probability is obtained for each
possible final state.

The above approach relies on an important property of the system which
is invariance of a reaction sequence with respect to changes of
initial particle spacings.  To understand this, consider again
Fig.~\ref{fig:traj}.  By altering the initial spacings of the
particles, the absolute times at which trajectories intersect and
reactions may occur (if the reactants have survived) may be altered.
For example, by increasing the spacing between the fifth and sixth
particles, the trajectories of the third and fourth particles can be
made to intersect first. However as we have already seen, for any
particle, the {\em order} of intersections it encounters does not
change and so the final states and probabilities are invariant. This
invariance is manifested in the matrix product by the fact that the
order in which we use the reduction rule (\ref{eqn:RL}) is unimportant
i.e.\ matrix multiplication is associative.  
%Thus, it is the
%invariance with respect to initial spacings that allows the system to
%be solved by using a product of matrices.

Averaging over all initial strings of length $N$ yields
\begin{equation}
F(N) = \langle W | (f_L L + f_R R)^N | V \rangle\;.
\label{eqn:F}
\end{equation}
To evaluate $F(N)$ we write $R = \sqrt{f_R^\ast f_L^\ast}/f_R \, a +
\eta_L$ and $L = \sqrt{f_R^\ast f_L^\ast}/f_L \, a^\dagger + \eta_R$.
One can check from (\ref{eqn:RL},\ref{eqn:VWdef}) that $a, a^\dagger$
satisfy a $q$-deformed harmonic oscillator algebra
\begin{eqnarray}
aa^\dagger - q a^\dagger a = 1-q
\label{eqn:ho} \\
  \langle W| a^\dagger =  \langle W|q/\sqrt{\chi} \qquad
  a |V\rangle = \sqrt{\chi}\, |V\rangle\;.
\label{eqn:coherent}
\end{eqnarray}
As is evident from (\ref{eqn:coherent}) the vectors $|W \rangle $, $|
V \rangle$ are eigenvectors of $a$, and are called $q$-deformed
coherent states. The explicit form of these eigenvectors is known
\cite{BECE}.  Using the above definitions (\ref{eqn:F}) becomes
\begin{equation}
F(N) = \langle W | \left[ 
 \sqrt{f_R^\ast f_L^\ast}\, ( a +a^\dagger) + 1{-}f_L^\ast{-}f_R^\ast
\right]^N | V \rangle.
\label{eqn:Fmat}
\end{equation}

In the deterministic limit, $q=0$, (\ref{eqn:ho}) becomes $a a^\dagger
=1$ and $a$, $a^\dagger$ are ladder operators. For $\chi = 1$, as in
\cite{EF}, one can see using (\ref{eqn:coherent}) that the matrix
product (\ref{eqn:Fmat}) is equivalent to a problem of counting 1D
random walks that do not return to the origin \cite{KS}.  For general
$q$, the evaluation of (\ref{eqn:Fmat}) poses a $q$-combinatoric
problem, the solution of which we now outline.

We take advantage of recent techniques and results \cite{BECE} for the
calculation of matrix products such as (\ref{eqn:Fmat}). The approach
is based on the fact that the eigenstates of the operator $x = a
+a^\dagger$ (analogous to the position operator in the usual harmonic
oscillator) can be expressed in terms of $q$-deformed Hermite
polynomials, whose orthogonality properties and generating functions
are known.  Decomposing $\langle W|$ and $| V\rangle$ onto the
eigenbasis of $x$ allows an integral representation of $F(N)$.  From
this expression the large $N$ behavior can be extracted by using
standard asymptotic analysis detailed in \cite{BECE}.  The results are
summarized in Table~II in which
\begin{equation}
 K= \prod_{n=1}^{\infty}
\frac{(1-q^n)^4}{(1-q^n \, \sqrt{\chi})^2 (1- q^n/\sqrt{\chi})^2}\;.
\label{eqn:Kdef}
\end{equation}
\begin{table}[htb]
\begin{center}
\begin{tabular}{c|c|c|c}
  &  $A$ & $B$ & $\gamma$\\[0.5ex]
\hline &&&\\
$ \chi < q^2$ &
$\displaystyle 1 - \frac{1}{q^2} \, \frac{f_R^\ast}{f_L^\ast}$ &
$ \left(1-q \right) \left(  f_L^\ast - \frac{f_R^\ast}{q}
 \right)$ &
$\displaystyle 0$
\\[2.9ex]
$ \chi = q^2 $ &
$\displaystyle \frac{1}{\sqrt{\pi}} \left( \frac{1}{f_L^\ast f_R^\ast}
  \right)^{\!\frac{1}{4}} $ &
$ \left( \sqrt{f_L^\ast} - \sqrt{f_R^\ast} \right)^{\!2}$ &
$\displaystyle \frac{1}{2}$
\\[2.9ex]
$  q^2 {<} \chi {<} 1 $ & 
$\displaystyle \frac{1}{\sqrt{4\pi}} \frac{K}{(1-\sqrt{\chi})^2}
 \left( \frac{1}{f_L^\ast f_R^\ast} \right)^{\!\frac{3}{4}}$ &
$ \left( \sqrt{f_L^\ast} - \sqrt{f_R^\ast} \right)^{\!2}$ &
$\displaystyle \frac{3}{2}$
\\[2.9ex]
$ \chi = 1 $ &
$\displaystyle \frac{1}{\sqrt{\pi}} \left( \frac{1}{f_L^\ast f_R^\ast}
  \right)^{\!\frac{1}{4}} $ &
$0$ &
$\displaystyle \frac{1}{2}$
\\[2.9ex]
$ \chi > 1$ & 
$\displaystyle \frac{1}{\sqrt{4\pi}} \frac{K}{(1-\sqrt{\chi})^2}
 \left( \frac{1}{f_L^\ast f_R^\ast} \right)^{\!\frac{3}{4}}$ & 
$ \left( \sqrt{f_R^\ast} - \sqrt{f_L^\ast} \right)^{\!2}$ & 
$\displaystyle \frac{3}{2}$
\\[2.9ex]
\end{tabular}
\end{center}
\caption{\label{table:survivalprobs}Explicit expressions of the parameters
  in the generic decay form $F(N)-F_\infty = A (1-B)^{N+\gamma} N^{-\gamma}$.
  For  $\chi\leq1,\; F_\infty=0$ 
and for $\chi > 1,\; F_\infty= 1 - f_L^\ast/f_R^\ast$. }
\end{table}
Using Eq.~\ref{eqn:Psum} and the results of Table~II, one can
calculate the asymptotic density (\ref{eqn:rhodef}) for any initial
spatial distribution of particles. In Table~I we present the results
found with $G(N;X)$ given by the Poisson initial condition.

In summary we have studied stochastic ballistic annihilation and
coalescence in 1D. The asymptotic density was calculated exactly for
Poisson initial conditions. The resulting phase diagram is described
by two parameters $q$ and $\chi$. The first is a measure of
stochasticity while the second encodes information about the reaction
processes and the initial densities.  The phase diagram contains two
new regimes which were not known before.

One application of our results is to generalize the surface growth
model of \cite{KS}.  In that model down (up) steps of a 1D interface
move deterministically to the right (left) and annihilate on meeting.
Thus the surface smoothens with time.  Our results for $\eta_L =
\eta_R = 0$ allow one to consider the effect of a probability $q$ of a
new terrace being formed when steps meet and the effect of an overall
tilted surface ($\chi \ne 1$).  We see that for $\chi=1$ the
probability $q$ does not affect the long-time smoothing of the
interface, whereas for a tilted interface the smoothing behaviour
changes for $q^2 \ge \chi$.

Our results are exact in the long-time limit.  It would be interesting
to study the approach to this asymptotic behavior.  One way to do this
would be through extensive numerical simulations although, as the
cross-over time is expected to grow with $q$, the simulation time
needed can be very large.  Previous numerical studies \cite{Sheu} were
performed at relatively short times and indicated that along the line
$\chi=1$ the decay exponent depends on $q$ which is not the case in
the true long-time limit.

Our results allow extension to other initial spatial distributions by
using the results in Table~II and appropriate forms for $G(N,X)$ in
(\ref{eqn:Psum}).  We expect other distributions, for which the number
of particles in a macroscopic region obeys a central limit theorem, to
exhibit the same phases. However, power law distributions might
generate different behavior.

Generalizations to higher dimensions and more than two velocities are
desirable.  Indeed, even in one dimension and with reaction
probability one, models with more than two velocities have been shown
to exhibit rich behavior \cite{3species}.  It would be interesting to
try and generalize the analytical approach presented here to that
case.

We acknowledge the EPSRC (R.A.B.) and the Israeli Science Foundation
(Y.K.) for financial support; we also thank D. Mukamel for useful
discussions.

\vspace{-0.5cm}
%%%%%%%%%%%%%%%%%%%%%%%%%%%%%%%%%%%%%%%%%%%%%%%%%%%%%%%%%%%%%%%%%%%%%%%%%%%%%%%
%                              Bibliography                                   %
%%%%%%%%%%%%%%%%%%%%%%%%%%%%%%%%%%%%%%%%%%%%%%%%%%%%%%%%%%%%%%%%%%%%%%%%%%%%%%%
%

\end{document}